\documentclass[cmp]{svjour}

\usepackage{latexsym}
\usepackage{graphics}
\usepackage{amsmath}
\usepackage{amssymb}

\def\stop{\hfill$\Box$}
\def\R{\mathbb{R}}
\def\Z{\mathbb{Z}}
\def\gi{g}
\def\go{\tilde{g}}
\def\lbar{\overline}
\def\Om{\Omega}
\def\S{\Sigma}
\def\tg{\tilde{g}}
\def\tM{\tilde{M}^3}
\def\del{\delta}
\def\norm{\parallel}
\def\ep{\epsilon}

\def\Mepm{\mathcal{M}_{M,\ep}}
\def\Mepo{\mathcal{M}_{\Omega, \ep}}
\def\Mepme{\mathcal{M}^G_{M, \ep}}
\def\Mepoe{\mathcal{M}^G_{\Omega, \ep}}
\def\Fep{F_{\ep}}
\def\Fepe{F^G_{\ep}}

\def\om{\omega}
\def\Dphi{D^{\prime}\Phi_o}
\def\Up{\Upsilon}
\def\vp{\varphi}
\def\na{\nabla} 
\def\12{\frac{1}{2}}
\def\hUp{\hat{\Up}}
\def\heta{\hat{\eta}}
\def\hvp{\hat{\vp}}
\def\The{\Theta}
\def\na{\nabla}
\def\nat{\nabla^t}
\def\nao{\nabla^0}
\def\gt{\tilde{g}(t)}
\def\g0{\tilde{g}(0)}
\def\ft{f(t)}
\def\pdi{\partial_i}
\def\pdj{\partial_j}
\def\pdk{\partial_k}

\def\pdn{\partial_n}
\def\pda{\partial_{\alpha}}
\def\pdb{\partial_{\beta}}
\def\pdd{\partial_{\delta}}
\def\ddt{\frac{d}{dt}}
\def\S{\Sigma}
\def\12{\frac{1}{2}}
\def\sim{\mathcal{S}}
\def\vnt{\vec{n}(t)}
\def\vno{\vec{n}(0)}
\def\vn{\vec{n}}
\def\a{\alpha}
\def\b{\beta}
\def\l{\lambda}
\def\d{\delta}
\def\Gam{\Gamma}
\def\Up{\Upsilon}
\def\vp{\varphi}
\def\oints{\oint_{\S}}
\def\cd{\cdot}
\def\R{\mathbb{R}}
\def\yml{Y^M_L}
\def\ep{\epsilon}
\def\pr{\prime}

\spnewtheorem*{mainthm}{Main Theorem}{\bf}{\it}
\spnewtheorem*{rmk}{Remark}{\it}{\rm}
\spnewtheorem*{rdlem}{Reduction Lemma}{\bf}{\it}
\spnewtheorem*{corofift}{Corollary of the IFT}{\bf}{\it}

\begin{document}

\title{On Existence of Static Metric Extensions in General Relativity}

\author{Pengzi Miao \inst{1} \fnmsep \inst{2}}

\institute{Department of Mathematics, Stanford University, CA 94305, USA. \\
E-mail: mpengzi@math.stanford.edu \\ \and
Mathematical Sciences Research Institute, Berkeley, CA 94720, USA. \\
E-mail: pengzim@msri.org}

\date{Received: 31 January 2003 / Accepted: 14 April 2003}

\communicated{G.W. Gibbons}

\maketitle

\begin{abstract}
Motivated by problems related to quasi-local mass in general relativity, we study the static metric extension conjecture proposed by R. Bartnik \cite{Bartnik_energy}. We show that, for any metric on $\bar{B}_1$ that is close enough to the Euclidean metric and has reflection invariant boundary data, there always exists an asymptotically flat and scalar flat {\em static} metric extension in $M = \R^3 \setminus B_1$ such that it satisfies Bartnik's geometric boundary condition \cite{Bartnik_energy} on $\partial B_1$.
\end{abstract}

\section{Introduction}
\label{intro}
Let $(M^3, g)$ be an asymptotically flat time-symmetric initial data set
satisfying conditions of the Positive Mass Theorem \cite{Schoen_Yau} in general relativity. It is an interesting and challenging question to ask how much energy or mass
can be localized in a bounded region $\Om \subset M^3$. The underlying idea is that we expect the total energy of a system can be consistently found by computing contributions from its separate components. Among various efforts towards understanding this question,  R. Bartnik gave his quasi-local mass definition $m_B(\Om)$ in \cite{Bartnik_localmass}, which seems to have many appealing properties. We recall that
$$ m_B(\Om) = \inf \{ m_{ADM}((\tilde{M}^3, \tilde{g})) \ | \ (\tilde{M}^3, \tilde{g}) \in \mathcal{PM} \}, $$
where $m_{ADM}(\cdot)$ is the ADM mass functional for asymptotically flat 
manifolds \cite{Bartnik_mass} and 
$\mathcal{PM}$ denotes the space of all 
$(\tilde{M}^3, \tilde{g})$ satisfying conditions of the Positive Mass 
Theorem, which contains $(\Om, g)$ isometrically and contains no horizon 
outside $\Om$. 
It is conjectured by R. Bartnik that there exists a $(\tilde{M}^3, \tilde{g}) \in \mathcal{PM}$, called a minimal mass extension, the mass of which realizes $m_B({\Om})$ and $\tilde{g}$ 
is a scalar flat and {\em static} metric outside $\Om$. 

In \cite{Corvino}, J. Corvino gave a detailed study of static metrics from a pure scalar curvature deformation point of view. He showed that, if a metric $g$ is not static in an open domain $U$, one can 
locally deform the scalar curvature of $g$ inside $U$. Corvino's result suggests an interesting proof of the second part of Bartnik's  conjecture on minimal mass extension, because if $(\tilde{M}^3, \tg)$ is such an extension  and $\tg$ is not static in $\tM \setminus \lbar{\Om}$, one can first bump the scalar curvature of $\tg$ up and then use conformal deformation to decrease the ADM mass of $(\tM, \tg)$.
  
Hence, the existence of a static metric $\tg$ outside $\Om$ satisfying some meaningful boundary condition on $\partial \Om$ that is relevant to the mass of $\tg$ becomes a basic question in understanding $m_B(\Om)$.
In \cite{Bartnik_energy}, R. Bartnik proposed the following extension problem with an interesting geometric boundary condition.

\vspace{.2cm}

\noindent \underline{Static Metric Extension Conjecture}: \\
\indent \emph{Given a bounded region $\Om \subset (M^3, g)$, there exists a scalar flat and static metric $\tg$ on $M \setminus \Om$ so that
$$ \hspace{2.7cm} g |_{\partial \Om} = \tg|_{\partial \Om} \ \ 
   \mathrm{and}\ \ H(\partial \Om, g) =  H(\partial \Om, \tg), 
   \hspace{2.3cm} ({\bf bd}) $$
where $ H(\partial \Om, g), H(\partial \Om, \tg)$ represents the mean curvature of $\partial
\Om$ with respect to $g, \tg$ following the unit normal vector pointing to the outside of $\Om$.} 

\vspace{.2cm}

For the motivation of the boundary condition ({\bf bd}) and its influence on 
the ADM mass of $(\tM, \tg)$, readers may refer to \cite{Miao_PMT} for a discussion.

In this paper, we study the above conjecture by taking $M = \R^3$ and $g$ to be a small perturbation of the Euclidean metric $g_o$. We first derive an analytical criteria that guarantees the existence of such an extension for a general domain $(\Om, g)$, then we focus on the case that $\Om$ is a round ball and  prove the following existence theorem.

\begin{mainthm}
Let $B_1$ be the unit open ball in $\R^3$. Then, for any number $\del \in (-1, -\frac{1}{2}]$ and any integer $k > \frac{9}{2}$, there exits a neiborhood $U$ of the Euclidean metric $g_o$ in $\mathcal{H}^k(B_1)$ such that, for any $g \in U$, if $g |_{S^2}$ and $H(S^2, g)$ is invariant under a $\Z_2 \times \Z_2 \times \Z_2$ action,
then there exists a scalar flat {\bf \em static} metric $\tg \in \mathcal{H}^k_{\del}(\R^3 \setminus B_1)$ such that
\begin{equation*} 
\left\{ \begin{array}{ccc}
        \go |_{S^2} & = & \gi |_{S^2} \\
        H(S^2, \go) & = & H(S^2, \gi) \ .
        \end{array}
\right. 
\end{equation*} 
Here $H(S^2, \gi)$, $H(S^2, \go)$ denotes the mean curvature of $S^2$ with respect to $\gi$, $\go$. The $\Z_2 \times \Z_2 \times \Z_2 $ invariance of $g|_{S^2}$ and $H(S^2, g)$ means that they are invariant under reflections about 
all the coordinate planes spanned by an arbitrary orthonormal basis $\{ e_1, e_2, e_3 \}$.
\end{mainthm}

\begin{rmk}
In fact, a slight modification of our argument shows that given any metric $\sigma$ and any function $h$ on $S^2$ that are sufficiently close to $g_o|_{S^2}$ and $H(S^2, g_o)$, if they satisfy the same symmetry condition as above, then there exists a scalar flat and static metric $\tg$ on $\R^3 \setminus B_1$ such that $g|_{S^2} = \sigma$ and $H(S^2, \tg)=h$.
\end{rmk}

\section{Preliminary}
\label{prelim}
We first recall the definition of a scalar flat metric being {\bf static}. 

\begin{definition}
Let $g$ be a metric with zero scalar curvature on an open set $U$. We say that $g$ is {\bf static} in $U$ if there exists a function $f$ on $U$ such that
\begin{equation} \label{staticeq}
\left\{
\begin{array}{rcl}
fRic(g) & = & Hess_{g}(f) \\ 
\triangle_{g}f & = & 0  \hspace{1.5cm} .
\end{array}
\right.
\end{equation} 
\end{definition}
We recommend \cite{Corvino} for a discussion of the origin of this definition and various properties of static metrics.

As in many problems involving small data, our main tool to obtain existence is the following corollary of the Implicit Function Theorem (See \cite{IFT}, \cite{Reula}).

\begin{corofift}
Let $X, Y, Z$ be Banach manifolds, $U, V$ be an open set of $X, Y$ and 
$G: V \times U \longrightarrow Z$ a differentiable function. Assume that there exists $x_o \in U, y_o \in V$ such that $G(y_o, x_o) = 0$, and $D^{\pr}G(y_o, x_o): Y \rightarrow Z$, the differential of $G$ with respect to the first argument is surjective and has complemented kernel. Then there exist a neighborhood $U_{x_o} \subset U, V_{y_o} \subset V$ around $x_o, y_o$ such that for any 
$x \in U_{x_o}$ there exists at least a $y \in V_{y_o}$ satisfying $G(y, x) = 0$.
\end{corofift}

We begin our investigation on Bartnik's conjecture by studying a general bounded domain $\Om \subset \R^3$ and its complement $M = \R^3 \setminus \Om$, where $\Om$ has a smooth boundary $\S$ such that $(\Om, \S)$ is diffeomorphic to $(B_1, S^2)$. By translation, we assume that $ 0 \in \Om$.  
For any $l \in \R$, we let
$H^l_{\Om}, H^l_{\S}$ denote the usual Sobolev space of functions 
on $\Om, \S$, and $\mathcal{H}^l_{\Om},\mathcal{H}^l_{\S}$ represent 
the space of $(0, 2)$ symmetric tensors on $\Om, \S$ whose components 
lie in $H^l_{\Om}, H^l_{\S}$. We define
$\mathcal{H}^l_{(\S, \R^3)}$ to be the space of $\R^3$-valued $1$-forms 
on $\S$, whose components lie in $H^l_{\S}$. 

For $\del \in \R$, $k$ a non-negative integer, we introduce $H^{k}_{\del, M}$, the weighted Sobolev space on $M$ with weight $\del$, following Bartnik's notation \cite{Bartnik_mass}:
\begin{equation}
H^{k}_{\del, M} = \{ u \in H^k_{loc}(M) \ | \ \norm u \norm_{k, 2, \del} < 
\infty \} \ , 
\end{equation}
where $H^k_{loc}(M)$ is the usual Sobolev space on $M$ and 
\begin{equation}
\norm u \norm_{k, 2, \del} = \sum^k_{l = 0} \left\{ 
\int_M \left( |D^l u| \cdot |x|^{l -\del} \right)^2 |x|^{-3} dx \right\}^
\frac{1}{2}.
\end{equation}
We then define $\mathcal{H}^k_{\del, M}$ to be the space of $(0, 2)$ symmetric tensors on $M$ whose components lie in $H^k_\del(M)$.

Given $\ep > 0$, we will work in the following spaces:
\begin{equation}
\begin{array}{lcl}
\mathcal{M}_{M, \ep} & = & \{ g_o + \The \ | \ \The = \The_{ij}dx^idx^j \in 
\mathcal{H}^k_{\del, M},  \ \norm \The_{ij} \norm_{k,2,\del} < \ep \} \\
\mathcal{M}_{\Om, \ep} & = & \{ g_o + \The \ | \ \The = \The_{ij}dx^idx^j \in \mathcal{H}^k_{\Om}, \ \norm \The_{ij} \norm_{H^k(\Om)} < \ep \} \\
F_\ep & =  & \{ 1 + \phi \ | \ \phi \in H^k_{\del, M}, \ \norm \phi \norm_{k,2,\del} < \ep \} 
\end{array}
\end{equation}
where we will always assume that $k - \frac{3}{2} > 3$ and
$\del < 0$. It follows from Sobolev imbeddings and weighted Sobolev inequalities \cite{Bartnik_mass} that we can choose $\ep$ sufficiently small so that $\Mepm, \Mepo$ only consists of $C^3$ metrics on $M, \Om$ and 
$\Fep$ only consists of $C^3$ positive functions on $M$. 

Throughout this paper we will use $\sim(\cdot)$ to denote the symmetrization operator on $(0, 2)$ tensors and use $\nabla_g(\cdot)$ to denote the covariant differentiation with respect to a metric $g$.
Our first lemma below shows that, to get a solution to the static equation (\ref{staticeq}) that is close to $g_o$, it suffices to consider a modified elliptic system. (See \cite{Reula} for a similar procedure.)

\begin{rdlem} \label{modifysys}
Let $\del \leq -\frac{1}{2}$. There exists a $\ep_0 > 0$ depending only on $\del$ such that if $\ep < \ep_0$ and $(\tg, f) \in \Mepm \times \Fep$ is a solution to 
\begin{equation}
\left\{ \begin{array}{rcc} \label{modifyeq} 
        fRic(\tg) - Hess_{\tg}(f) & = & f \mathcal{S}(\nabla_{\tg} \omega) \\
        \triangle_{\tg}(f) & =& 0 
        \end{array} 
\right.
\end{equation}
in $M$ with $\om = 0 $ on $\S$, where $\omega = \omega(g_o, \tg)$ is a $1$-form defined by
\begin{equation} \label{defofom}
\omega = div_{g_o}\tg - \frac{1}{2}d(tr_{g_o}\tg)  \ ,
\end{equation}
then $\om$ vanishes identically in $M$ and hence $(\tg, f)$ is a solution to $\mathrm{(\ref{staticeq})}$.
\end{rdlem}

\begin{proof}
We let `` ; '' denote covariant differentiation with respect to $\tg$ in 
local coordinates. Taking $\tg$-trace, $\tg$-divergence of (\ref{modifyeq}) and applying 
the contracted second Bianchi identity, we have that 
\begin{equation} \label{divtr}
\left\{ \begin{array}{ccl} 
         dR(\tg)_i & = & (\triangle_{\tg}\om)_i + {\tg}^{jk}\om_{j;ik} 
                      + 2\mathcal{S}(\nabla_{\tg}\om)(\frac{1}{f}\nabla_{\tg}f, \partial_i) \\
         R(\tg) & =  & {\tg}^{jk}\om_{j;k} \ .
        \end{array}
\right. 
\end{equation}
It follows from (\ref{divtr}), the Ricci identity and the boundary assumption that
\begin{equation} \label{eqofom}
\left\{
\begin{array}{cccc}
(\triangle_{\tg} \om)  + 2\mathcal{S}(\nabla_{\tg}\om)(\frac{1}{f}\nabla_{\tg}f, \cdot) + 
Ric(\tg)(\om, \cdot) & = & 0 & \  in\  M \\ 
\om & = & 0 & \ on\ \S
\end{array}
\right.
\end{equation}
where $\triangle_{\tg} \om$ denotes the rough Laplacian of the 
$1$-form $\om$ and $\{ \partial_1, \partial_2, \partial_3 \}$ is a standard
basis for $(\R^3, g_o)$.
We note that $(\ref{eqofom})$ is a second order linear elliptic system
of $\om_i \in H^{k-1}_{\del-1}$. When $(\tg, f) = (g_o, 1)$, it reduces to 
\begin{equation} \label{flateq}
\left\{ \begin{array}{cccl} 
         \triangle_{g_o} \om & = & 0 & \ \ \ in\ M \\
         \om & = & 0 & \ \ \ on\ \S \ .
        \end{array}
\right.
\end{equation}
Integrating by parts and using the decay assumption $\del \leq -\frac{1}{2}$, we see that (\ref{flateq}) only admits zero 
solution in $H^{k-1}_{\del - 1}$. Since injectivity 
is a continuous property for elliptic operators, we know that there exists
a $\ep_0 > 0$ so that if $\norm \tg - g_o \norm_{k,p,\del} < \ep_0$ and $\norm
f -1 \norm_{k,p, \del} < \ep_0$, $(\ref{eqofom})$ only admits zero 
solution as (\ref{flateq}) does. Hence, $\om$ vanishes identically in $M$ and $(\tg, f)$ solves (\ref{staticeq}).  
\end{proof}

\section{Linearization at the Flat Metric }
\label{lineariz}
From now on, we assume that $\del \leq -\frac{1}{2}$ and $\ep < \ep_0$.
Our Reduction Lemma suggests the following map between two Banach manifolds
$$ \Phi: \mathcal{M}_{M,\ep} \times F_{\ep} \times \mathcal{M}_{\Om,\ep} 
\longrightarrow 
\mathcal{H}^{k-2}_{\del-2, M} \times H^{k-2}_{\del-2, M} \times \mathcal{H}^{k - \frac{3}{2}}_{(\S, \R^3)} \times \mathcal{H}^{k - \frac{1}{2}}_{\S} \times
   H^{k - \frac{3}{2}}_{\S} $$ 
\begin{equation}
\Phi(\go, f, \gi) = \left(
\begin{array}{c}
fRic(\go) - Hess_{\go}f - f \mathcal{S}(\nabla_{\go}\om) \\
\triangle_{\go} f \\
\omega \\
\go |_{\S} - \gi |_{\S} \\
H(\S, \go) - H(\S, \gi)
\end{array}
\right),
\end{equation}
where $\om$ is defined by (\ref{defofom}).
It is readily seen that $\Phi$ is a differentiable map and 
$\Phi(g_o, 1, g_o) = 0$. Hence, to apply the Implicit Function Theorem, it is necessary to study
$$ \Dphi:
\mathcal{H}^k_{\del, M} \times H^{k}_{\del, M} 
\longrightarrow
\mathcal{H}^{k-2}_{\del-2, M} \times H^{k-2}_{\del-2, M} \times \mathcal{H}^{k - \frac{3}{2}}_{(\S, \R^3)} \times \mathcal{H}^{k - \frac{1}{2}}_{\S} \times
   H^{k - \frac{3}{2}}_{\S} $$
where $\Dphi$ denotes the partial differential
of $\Phi$ at $(g_o, 1, g_o)$ with respect to the first two arguments. 

\begin{lemma} \label{linearization}
Given $(\The, \phi) \in \mathcal{H}^k_{\del, M} \times H^{k}_{\del, M} $, we have that 
\begin{equation}
\Dphi(\The, \phi) = \left(
\begin{array}{c}
-\frac{1}{2}\triangle \The - Hess \phi \\
\triangle \phi \\
div \The - \frac{1}{2}d tr \The \\
\The |_{\S} \\
-\frac{1}{2}\The_{nn;n} + \frac{1}{2}H_o \The_{nn} - < \The |_{\S}, \Pi_o> +
(div \The - \frac{1}{2}d tr \The)_n 
\end{array}
\right)
\end{equation}
where ``$\triangle(\cdot), div(\cdot), tr(\cdot)$'' each is taken with respect 
to $g_o$, $\Pi_o$ is the second fundamental form of $\S$ in $\R^3$, $H_o$ is the mean curvature of $\S$ in $\R^3$,
a tensor with a lower index ``$n$'' denotes its value evaluated at the normal vector 
$\nu$ on $\S$ pointing to $\infty$ and ``;'' denotes the covariant differentiation with respect to $g_o$ in local coordinates.
\end{lemma}

\begin{proof}
Let $\{( \gt, \ft) \}_{|t|< 1}$ be a family of metrics and functions on $(M, \S)$ such that $(\g0, f(0)) = (g_o, 1)$ and $({\tg}^\prime(0), f^\prime(0)) = (\The, \phi)$. 
We view $g_o$ as a background metric. 
For each $t$, we let $D^t$ denote the connection determined by $\gt$ and 
$\na^t(\cdot)$ denote the covariant differentiation with respect to $\gt$. 
We also let $R(t)_{ij} dx^i dx^j$ denote the Ricci tensor of $\gt$.

Since the difference between any two connections is a tensor, we can write
\begin{equation}
 D^t_{\pdi}\pdj - D^0_{\pdi}{\pdj} = D(t)^{k}_{ij}\pdk \ ,
\end{equation}
where
\begin{equation} \label{Dij}
D(t)^k_{ij} = \frac{1}{2} \gt^{kl} \{ \gt_{lj;i} + \gt_{li;j} - \gt_{ij;l} 
\} \ .
\end{equation}
It follows from the definition of the Ricci tensor that
\begin{equation}
R(t)_{ij} - R(0)_{ij} = D(t)^k_{ij;k} - D(t)^k_{ki;j} + D(t)^k_{kl}D(t)^l_{ij} - D(t)^k_{lj}D(t)^l_{ik} \ ,
\end{equation}
which gives that
\begin{equation} \label{drgt}
\ddt{R(t)_{ij}}|_{t=0} = D^\prime(0)^k_{ij;k} - D^\prime(0)^k_{ki;j} \ .
\end{equation} 
On the other hand, we know from (\ref{Dij}) and the fact $\g0_{ij;k}=0$ that
\begin{equation} \label{Dijks0}
D^\prime(0)^k_{ij;s}  = \frac{1}{2} \g0^{kl} \{ \The_{lj;is} +
\The_{li;js} - \The_{ij;ls} \} \ .
\end{equation}
Hence, (\ref{drgt}), (\ref{Dijks0}) and the fact that $\g0$ is flat imply that
\begin{eqnarray} \label{caldrgt}
\ddt{R(t)_{ij}}|_{t=0} 
& = & -\frac{1}{2}(\triangle \The)_{ij} -\frac{1}{2}Hess(tr \The)_{ij}
+ \12 \g0^{kl} \{ \The_{lj;ki} + \The_{li;kj} \} \nonumber \\
&  & + \12 \g0^{kl} \{ \The_{lj;ik} - \The_{lj;ki} + 
\The_{li;jk} -\The_{li;kj} \}  \nonumber \\
& = &  -\frac{1}{2}(\triangle \The)_{ij}
-\frac{1}{2}Hess(tr \The)_{ij}  +  \mathcal{S}(\nao div \The) _{ij} \ .
\end{eqnarray}
We rewrite (\ref{caldrgt}) as
\begin{equation} \label{linricci}
\ddt{R(t)_{ij}}|_{t=0} = -\frac{1}{2}(\triangle \The)_{ij} + \mathcal{S} [\nao (div \The - \12 d(tr \The)) ]_{ij} \ .
\end{equation}
To identify the non-elliptic term in (\ref{linricci}), we compute $\ddt \sim(\nat div \gt ) |_{t=0}$ and $ \ddt Hess_{\gt}(tr \gt) |_{t=0}  $. First, we note that
\begin{equation}
(\nat div \gt)_{ij} - (\nao div \gt)_{ij} = - D(t)^k_{ij}(div \gt )_k \ .
\end{equation}
Hence, 
\begin{eqnarray}
\ddt \sim(\nat div \gt )_{ij} |_{t=0} & = & \ddt \sim(\nao div \gt )_{ij} |_{t=0} \\
& = & \sim (\nao div \The )_{ij} \ .
\end{eqnarray}
Second, by definition we have that
\begin{equation}
(Hess_{\gt} tr \gt )_{ij} =   \pdi \pdj tr \gt - (D^0_{\pdi} \pdj) tr \gt 
 - (D(t)^k_{ij}\pdk) tr \gt \ ,
\end{equation}
which implies that
\begin{equation} 
\ddt (Hess_{\gt} tr \gt )_{ij} |_{t=0} = \pdi \pdj tr \The - (D^0_{\pdi}
\pdj)
tr \The
=  (Hess tr \The )_{ij} 
\end{equation}
because $tr g(0)$ is a constant. Thus we have that
\begin{equation}
\ddt \{ R(t)_{ij} - \sim ( \nat \omega(t) )_{ij}  \} |_{t=0}  =
-\frac{1}{2}(\triangle \The)_{ij} \ ,
\end{equation}
where $\omega(t)$ is given by
\begin{equation}
\omega(t) = div \gt - \12 d (tr \gt) \ .
\end{equation}
A similar calculation gives that
\begin{equation} \label{linhess}
\ddt (Hess_{\gt} \ft )_{ij} |_{t=0} = \pdi \pdj \phi - (\nao_{\pdi} \pdj)\phi
=  (Hess \phi )_{ij} 
\end{equation}
and
\begin{equation}
\ddt \{ \triangle_{\gt} \ft \} |_{t=0} = \triangle \phi \ .
\end{equation}

Next we proceed to linearize the mean curvature functional at $\S$. We define $\vnt$ to be the outward unit normal vector field to $\partial \Omega$ determined by $\gt$. We also choose $\{ x^1, x^2, x^3 \}$ to be a local coordinate chart for $M$ such that $\{x^1, x^2\}$ gives a local chart for $\S$ and $\partial_3$ coincides with $\vn(0) = \vn$. 
Then by definition, 
\begin{equation} \label{defH}
H(\S, \gt)(x) = \gt^{\a \b} \Pi(t)_{\a \b} = \gt^{\a \b} <D^t_{\pda}\pdb, \vnt>_{\gt} \ ,
\end{equation}
where $\Pi(t)$ is the second fundamental form of $\S$ with respect to $\gt$. Henceforth, we let $\a, \b, \ldots$ run through $\{ 1, 2 \}$ and $i, j, \ldots$ run through $\{1, 2, 3 \}$. We will also use the lower index `` $n$ '' to specially denote the index ``$3$''. 
 
It follows from (\ref{defH}) that
\begin{eqnarray} \label{f}
\ddt H(\S, \gt)(x) |_{t=0} & = & {\tg}^\prime(0)^{\a\b}<D^0_{\pda}\pdb, \vno>_{g_o}
 \nonumber \\
&  & + g_o^{\a \b} \ddt \{<D^t_{\pda}\pdb, \vnt>_{\gt} \} |_{t=0} \ , 
\end{eqnarray}
where
\begin{equation} \label{g}
{\tg}^\prime(0)^{\a\b}<D^0_{\pda}\pdb, \vnt>_{g_o} = 
-< \The |_{\S}, \Pi_o >_{g_o} \ , 
\end{equation}
and
\begin{eqnarray}
\ddt \{<D^t_{\pda}\pdb, \vnt>_{\gt} \} |_{t=0} & = &
\The(D^0_{\pda} \pdb, \vno) + <D^0_{\pda} \pdb, \vec{n}^\prime(0)>_{g_o}
\nonumber \\
& & + < \ddt \{ D^t_{\pda} \pdb \}|_{t=0}, \vno>_{g_o} \ .
\end{eqnarray}
Straightforward calculation gives that
\begin{equation} \label{a}
\The(D^0_{\pda} \pdb, \vno) = \The(\Gam^{n}_{\a \b} \pdn + \Gam^{\d}_{\a
\b}
\pdd, \pdn) = \The_{nn}\Gam^{n}_{\a \b} + \Gam^{\d}_{\a \b} \The_{\d n} \ , 
\end{equation}
where $\Gam^{k}_{ij}$ denotes the Christoffel symbols for $g_o$, and
\begin{equation} \label{3}
<D^0_{\pda} \pdb, \vec{n}^\prime(0)>_{g_o} = <\Gam^{n}_{\a \b} \pdn +
\Gam^{\d}_{\a \b} \pdd,  \vec{n}^\prime(0)>_{g_o} \ .
\end{equation}
On the other hand, the fact that $<\vnt, \pdd>_{\gt} = 0$ and $<\vnt, \vnt>_{\gt} = 1$ imply that
\begin{equation}
\left\{
\begin{array}{rcl}
\The(\pdn, \pdd) + <\pdd, \vec{n}^\prime(0)>_{g_o} &  = & 0 \\
\The(\pdn, \pdn) + 2<\vec{n}^\prime(0), \vno>_{g_o} &  = & 0 \ .
\end{array}
\right.
\end{equation}
Hence, (\ref{3}) becomes 
\begin{equation} \label{b}
<D^0_{\pda} \pdb, \vec{n}^\prime(0)>_{g_o} = -\12 \The_{nn}\Gam^n_{\a \b} - 
\The_{n \d} \Gam^{\d}_{\a \b} \ . 
\end{equation}
To calculate  $< \ddt \{ D^t_{\pda} \pdb \}|_{t=0}, \vno>_{g_o}$, we recall 
that 
$$ D^t_{\pda } \pdb - D^0_{\pda} \pdb = D(0)^k_{\a \b} \pdk \ .$$ 
Hence, 
\begin{equation} \label{c}
< \ddt \{ D^t_{\pda} \pdb \}|_{t=0}, \vno>_{g_o}  =  D^\prime(0)^{n}_{\a \b}
\ ,
\end{equation}
where
\begin{equation}
D^\prime(0)^{n}_{\a \b} = \frac{1}{2} g_o^{nn} \{ \The_{n \a; \b} + \The_{n \b;\a}  - \The_{\a \b; n} \}
\end{equation}
by (\ref{Dij}). Therefore (\ref{a}), (\ref{b}) and (\ref{c}) imply that
\begin{equation} \label{d}
g_o^{\a \b} \ddt \{<D^t_{\pda}\pdb, \vnt>_{\gt} \} |_{t=0} =   \12 H(\S, g_o) \The_{nn} + g_o^{\a \b} D^\prime(0)^{n}_{\a \b} \ .
\end{equation}
To see the geometric meaning of the second term in (\ref{d}), we compute 
\begin{equation}
(div \The)_n = g_o^{ij}\The_{ni;j}= \The_{nn;n} + g_o^{\a \b} 
\The_{n \a; \b} 
\end{equation}
and 
\begin{equation}
(d tr \The)_n = \The_{nn;n} + g_o^{\a \b} \The_{\a \b;n} \ ,
\end{equation}
which imply that
\begin{equation} \label{e}
 g_o^{\a \b} \The_{n \a; \b} = (div \The - \12 d tr \The)_n -\12 \The_{nn;n}\ .
\end{equation}
Therefore, it follows from (\ref{f}), (\ref{g}), (\ref{d}) and (\ref{e}) that
\begin{eqnarray}
\ddt H(\S, \gt)(x) |_{t=0} &  = & -\12 \The_{nn;n} + \12 H(\S, g_o) \The_{nn} 
\nonumber \\
& &  -< \The |_{\S}, \Pi_o >_{g_o} +  (div \The - \12 d tr \The)_n \ ,
\end{eqnarray} 
which proves the lemma. 
\end{proof}

\section{Derivation of Potential Obstruction}
\label{obs}
Simple observation reveals that $\Dphi$ is equivalent to 
another operator $T$ that has a simpler boundary map
$$ T: \mathcal{H}^k_{\del, M} \times H^{k}_{\del, M} 
\longrightarrow 
\mathcal{H}^{k-2}_{\del-2, M} \times H^{k-2}_{\del-2, M} \times \mathcal{H}^{k - \frac{3}{2}}_{(\S, \R^3)} \times \mathcal{H}^{k - \frac{1}{2}}_{\S} \times
   H^{k - \frac{3}{2}}_{\S} $$ 
\begin{equation}
T(\The, \phi) = \left(
\begin{array}{c}
-\frac{1}{2}\triangle \The - Hess\phi \\
\triangle \phi \\
div \The - \frac{1}{2}d tr \The \\
\The |_{\S} \\
-\frac{1}{2}\The_{nn;n} + \frac{1}{2}H_o \The_{nn} 
\end{array}
\right).
\end{equation}
Assuming that $\del$ is a non-exceptional value \cite{Bartnik_mass}, i.e. $\del \notin \mathbb{Z}$ in our case, we have the following important fact,

\vspace{.2cm}
\noindent \underline{Fact}:   
\emph{$T$ is an elliptic operator in the sense of $H\ddot{o}rmander$ \cite{Hormander} which includes the $Lopatinski\check{i}$-$\check{S}apiro$ conditions for the boundary map. 
Hence, $T$ is Fredholm and its image is determined by Coker(T), the kernel of its adjoint.}

\begin{rmk}   
In general it is a subtle problem to give a boundary condition for the Ricci curvature tensor such that it is both elliptic and geometric.  
Hence we have a non-trivial fact that ({\bf bd}) is an elliptic condition for the static metric equation. 
We omit its proof here since it is straightforward checking against the definition.
\end{rmk}

\begin{lemma} \label{obstruction}
For $ -\frac{3}{2} < \del \leq -\frac{1}{2}$ and $\del \neq -1$, $(\Up, \vp, \eta, \tau, h) \in Coker(T)$ if and only if 
\begin{equation}
\hspace{1.2cm}
\left\{ \begin{array}{ccc}
         \triangle \Up & = & 0 \\
         \triangle \vp & = & 0 
        \end{array}
\right. \hspace{2cm} in \ M 
\end{equation}
and 
\begin{equation} \label{simbdeqofad}
\left\{ \begin{array}{ccc}  
         \vp - \Up_{nn} & = & 0 \\
         \frac{\partial \vp}{\partial \vn } - div_{\S}\Up(\vn, \cdot) & = & 0 \\
         div \Up & = & 0 \\
         \Up |_{\S} & = & w g_o |_{\S} 
\end{array}
\right. \ \ \ on \ \S 
\end{equation}
where $\vn$ is the outward unit normal vector field to $\S$, $\Up(\vn, \cdot)$ is viewed as a $1$-form defined on $\S$ and $div_{\S}(\cdot)$ represents the divergence operator on $(\S, g_o|_{\S})$. 
\end{lemma} 

\begin{proof}
It follows from the general elliptic theory \cite{McOwen}, \cite{bvp} that $(\Up, \vp, \eta, \tau, h) \in Coker(T)$ if and only if
\begin{eqnarray} \label{def}
0 & = & \int_M <-\frac{1}{2}\triangle\The - Hess\phi, \Up> + \int_M \triangle \phi \cdot \vp  + \oint_{\S} <div \The - \frac{1}{2}d tr \The , \eta> \nonumber \\ 
& &  + \oint_{\S} <\The |_{\S} , \tau> + \oint_{\S} \{ -\frac{1}{2}\The_{nn;n} + \frac{1}{2}H_o \The_{nn} \} \cdot h   
\end{eqnarray}
for any $(\The, \phi) \in \mathcal{H}^{k}_{\d, M} \times H^k_{\d, M}$, where all the inner product between tensors are taken with respect to $g_o$.
Integrating by parts, we have that
\begin{eqnarray} \label{interior}
& &  \int_M <-\frac{1}{2}\triangle\The - Hess\phi, \Up> + \int_M \triangle \phi \cdot \vp    \nonumber \\
& =  & -\frac{1}{2} \int_M < \The, \triangle \Up > + \12 \oints < \na_{n} \The, \Up> - \12 \oints < \The, \na_n \Up> \nonumber \\
& & - \int_M div(div \Up) \cdot \phi + \oints \Up(\na \phi, \vn> - \oints (div \Up) (\vn) \cdot \phi \nonumber \\
& & + \int_M \phi \cdot \triangle \vp - \oints \vp \cdot \frac{\partial \phi}{\partial \vn} + \oints \phi \cdot \frac{\partial \vp}{\partial \vn} \ ,
\end{eqnarray}
where $\na_n (\cdot)$ represents the covariant derivative of a tensor along $\vn$ and $\na f$ denotes the $g_o$-gradient of a function $f$. Since $(\The, \phi)$ can be arbitrary, we have that 
\begin{equation} \label{inteq}
\left\{ \begin{array}{cll}
\triangle \vp - div(div \Up) & = & 0 \\
\triangle \Up & = & 0 
\end{array}
\right. \hspace{1cm} in \ M \ . 
\end{equation}

Now we begin to work in Gaussian coordinate chart $\{ x^1, x^2, x^3 \}$ 
around $\S$ in which 
$$ g_o = (dx^3)^2 + g_o(x)_{\a \b} dx^{\a} dx^{\b}  $$
and $\partial_ {x^3}$ coincides with $\vn$ along $\S$. 
Inside such a chart, we let `` ; '' denote the covariant differentiation with respect to $g_o$ and `` , '' denote the usual partial derivative. 
On $\S$, it follows from (\ref{def}), (\ref{interior}) and (\ref{inteq}) that 
\begin{eqnarray} \label{bdry}
0 & = & \oint_{\S} <div \The - \12 dtr\The, \eta>
 + \oint_{\S} \{ -\frac{1}{2}\The_{nn;n} 
+ \frac{1}{2}H_o \The_{nn} \} \cdot h  \nonumber \\ 
& & + \oint_{\S} <\The |_{\S} , \tau> + \12 \oints < \na_{n} \The, \Up> - \12 \oints < \The, \na_n \Up>  
 \nonumber \\ 
& & + \oints \Up(\na \phi, \vn>  - \oints (div \Up) (\vn) \cdot \phi
- \oints \vp \cdot \frac{\partial \phi}{\partial \vn} + \oints \phi \cdot \frac{\partial \vp}{\partial \vn}  \ .
\end{eqnarray} 
Integrating by parts over $\S$ and using the fact $\Up(\na \phi, \vn) = \Up_{nn} \frac{\partial \phi}{\partial \vn} + \Up(\vn, \na_{\S} \phi ) $, we have that
\begin{equation}
\oints \Up(\na \phi, \vn) = \oints \Up_{nn} \frac{\partial \phi}{\partial \vn}
- \oints div_{\S} [\Up(\vn, \cdot)] \phi \ .
\end{equation}
Since $\phi$ and 
$\frac{\partial \phi}{\partial \vn}$ 
can be independently chosen arbitrary, (\ref{bdry}) implies that 
\begin{equation}
\left\{ \begin{array}{cll}
\frac{\partial \vp}{\partial \vn} -  (div \Up)(\vn) -  div_{\S} [\Up(\vn, \cdot)] & = & 0 \\
 \vp  - \Up_{nn}  & = & 0 \ 
\end{array}
\right. \hspace{.3cm} on \ \S \ ,
\end{equation}
and (\ref{bdry}) is reduced to 
\begin{eqnarray} \label{bdrysim}
0 & = & \oint_{\S} <div \The - \12 dtr\The, \eta>
 + \oint_{\S} \{ -\frac{1}{2}\The_{nn;n} 
+ \frac{1}{2}H_o \The_{nn} \} \cdot h  \nonumber \\ 
& & + \oint_{\S} <\The |_{\S} , \tau> + \12 \oints < \na_{n} \The, \Up> - \12 \oints < \The, \na_n \Up>    \ .
\end{eqnarray}
 
To see the hidden relation among $\{ \Up, \eta, \tau, h \} $ on $\S$, we need to rewrite every integral in (\ref{bdrysim}) in terms of the independent free boundary quantities 
\begin{equation} \label{freedata}
\{\The |_{\S}, \The_{n \a}, \The_{nn}, (\na_n \The) |_{\S}, (\na_n \The)_{n \a}, (\na_n \The)_{nn}\} \ . 
\end{equation}
First, we have that
\begin{eqnarray} \label{nanthe}
 \oints < \na_{n} \The, \Up> & = & \oints < (\na_n \The) |_{\S}, \Up |_{\S}> + 
\oints (\na_n \The)_{nn} \cdot \Up_{nn} \nonumber \\
& & + 2 \oints < (\na_n \The)(\vn, \cdot), \Up(\vn, \cdot)> \  
\end{eqnarray}
\begin{eqnarray} \label{nanup}
 \oints < \na_{n} \Up, \The> & = & \oints < (\na_n \Up) |_{\S}, \The |_{\S}> + 
\oints (\na_n \Up)_{nn} \cdot \The_{nn} \nonumber \\
& & + 2 \oints < (\na_n \Up)(\vn, \cdot), \The(\vn, \cdot)> \ , 
\end{eqnarray}
where $(\na_n \Up)(\vn, \cdot), \The(\vn, \cdot), (\na_n \The)(\vn, \cdot), \Up(\vn, \cdot) $ each is treated as a 1-form on $\S$.
Second, we have that
\begin{equation}
\oints <div \The, \eta> = \oints (div \The)_n \cdot \eta_n + \oints (div \The)_{\a} \cd  \eta_{\b} \cd  g_o^{\a \b} 
\end{equation} 
\begin{equation}
\oints < dtr\The, \eta>  =  \oints (dtr \The)_n \cd \eta_n + \oints (dtr \The)_{\a} \cd  \eta_{\b} \cd  g_o^{\a \b} \ ,
\end{equation}
where 
\begin{equation} \label{divn0}
(div \The)_n = g_o^{ij} \The_{in;j} = g_o^{nn} \The_{nn;n} + g_o^{\a \b} \The_{\a n; \b} \ .
\end{equation}
To calculate $\The_{\a n; \b} $, we note that
\begin{equation} \label{m}
\left\{
\begin{array}{ccc}
\The_{\a n; \b} & = & \The_{\a n, \b} - \The_{\a i} \Gam^i_{n \b} - \The_{i n} \Gam^i_{\a \b} \\
div_{\S} [\The(\vn, \cd)] & = & g_o^{\a \b} \The_{n \a;^{\S} \b} \\ 
\The_{n \a;^{\S} \b} & = & \The_{\a n, \b} - \The_{\d n} \Gam^{\S \d}_{ \a \b} \ ,
\end{array}
\right.
\end{equation}
where `` $;^{\S}$ '' and $\Gam^{\S \d}_{\a \b}$ denote the covariant differentiation and the Christoffel symbol of the induced metric $g_o |_{\S}$ on $\S$.
It follows from (\ref{m}) and the fact $\Gam^{\S \d}_{ \a \b} = \Gam^{\d}_{\a \b}$ that
\begin{equation}
\The_{\a n; \b} =\The_{n \a;^{\S} \b} - \The_{nn}\Gam^n_{\a \b} -  \The_{\a \d} \Gam^{\d}_{n \b} - \The_{\a n} \Gam^n_{n \b} \ , 
\end{equation}
which implies that 
\begin{equation}
g^{\a \b} \The_{\a n; \b}  = 
div_{\S} [\The(\vn, \cd)] - H_o \The_{nn} + < \The |_{\S}, \Pi_o>
\end{equation}
by the fact that $\Gam^{\d}_{\b n} = - (\Pi_o)_{\l \b} g^{\l \d}$ and $\Gam^n_{\b n} = 0$.  
Therefore, (\ref{divn0}) becomes
\begin{equation} \label{divn}
(div \The)_n = div_{\S} [\The(\vn, \cd)] + \The_{nn;n} - H_o \The_{nn} + < \The |_{\S}, \Pi_o> \ .
\end{equation}
Next we calculate $(div \The)_{\d}$ and $(div_{\S} [\The |_{\S}] ) _{\d}$.
By definition, 
\begin{equation}
\left\{
\begin{array}{ccc}
(div \The)_{\d} &  = & \The_{n \d; n} + g_o^{\a \b}\The_{\a \d ; \b} \\
(div_{\S} [\The |_{\S}] ) _{\d} &  = &  g_o^{\a \b} \The_{\a \d ;^{\S} \b}
\end{array}
\right.
\end{equation}
where
\begin{eqnarray}
g_o^{\a \b} \The_{\a \d ; \b} & = & g_o^{\a \b} \{ \The_{\a \d , \b} - \The_{i \d}\Gam^i_{\a \b} - \The_{\a i}\Gam^i_{\d \b} \}  \nonumber \\
& = & (div_{\S} [\The |_{\S}] ) _{\d} - H_o \The_{n \d} - g_o^{\a \b} \The_{\a n}(\Pi_o)_{\d \b} \ .  
\end{eqnarray}
Hence, we have that
\begin{equation} \label{divd}
(div \The)_{\d} =  \The_{n \d; n} + (div_{\S} [\The |_{\S}] ) _{\d} - H_o \The_{n \d} - g_o^{\a \b} \The_{\a n}(\Pi_o)_{\d \b} \ .
\end{equation}
Similar calculations shows that
\begin{eqnarray}
(d tr \The)_{\a} & = & \The_{nn ; \a} + g_o^{\b \d} \The_{\b \d ; \a} 
\nonumber \\
& = & \The_{nn , \a} - 2 \The_{ni} \Gam^i_{n \a} + g_o^{\b \d} \{ \The_{\b \d , \a} - \The_{i \d} \Gam^i_{\b \a} - \The_{\b i}\Gam^i_{\d \a} \} \nonumber \\
& = & \The_{nn , \a} + 2 \The_{n \b} (\Pi_o)_{\d \a}g_o^{\d \b} + g_o^{\b \d} \The_{\b \d ;^{\S} \a } - 2g_o^{\b \d} \The_{n \d} (\Pi_o)_{\b \a} \nonumber \\
& = &  \The_{nn , \a} +  g_o^{\b \d} \The_{\b \d ;^{\S} \a } \ .
\end{eqnarray}
Therefore, integrating by parts on $\S$, we have that
\begin{eqnarray}
&  & \oints <div \The - \12 dtr \The, \eta > \nonumber \\
& = & - \oints <\The(\vn, \cd), d_{\S}\eta_n>  + \12 \oints \The_{nn;n} \cd \eta_n + \oints < (\na_n \The)(\vn, \cd),  \eta |_{\S} >  \nonumber \\
& & -  \oints H_o \The_{nn} \cd \eta_n  + \oints < \The |_{\S}, \eta_n \Pi_o>  
- \oints <(\The |_{\S} ,  \mathcal{S} [ \na_{\S} (\eta |_{\S})] >  \nonumber \\ 
& &  - \oints H_o < \The(\vn, \cd),  \eta |_{\S} >  - \oints < \The(\vn, \cd), \Pi_o((\eta |_{\S})^*, \cd)> \nonumber \\
& &  - \12 \oints  <(\na_n \The) |_{\S},  \eta_n \cd g_o |_{\S}>   +\12 \oints \The_{nn} \cd div_{\S} (\eta |_{\S}) \nonumber \\
& &  + \12 \oints <\The |_{\S}, div_{\S} (\eta |_{\S}) \cd  g_o |_{\S}>  \ ,
\end{eqnarray}
where $d_{\S}(\cdot)$ denotes the exterior derivative on $\S$ and $(\eta |_{\S})^*$ denotes the tangent vector on $\S$ that is the dual of $\eta |_{\S}$ with respect to $g_o |_{\S}$. 
Now we are in a position to rewrite (\ref{bdrysim}) as
\begin{eqnarray} \label{bdrysim2}
0 & = &  - \oints <\The(\vn, \cd), d_{\S}\eta_n>  + \12 \oints \The_{nn;n} \cd \eta_n + \oints < (\na_n \The)(\vn, \cd),  \eta |_{\S} >  \nonumber \\
&  & -  \oints H_o \The_{nn} \cd \eta_n  + \oints < \The |_{\S}, \eta_n \Pi_o> - \oints <(\The |_{\S},  \mathcal{S} [\na_\S (\eta |_{\S})] >  \nonumber \\ 
&  &  - \oints H_o < \The(\vn, \cd),  \eta |_{\S} >  - \oints < \The(\vn, \cd), \Pi_o( (\eta |_{\S})^*, \cd)> \nonumber \\
&  &  - \12 \oints  <(\na_n \The) |_{\S},  \eta_n \cd g_o|_{\S}>   +\12 \oints \The_{nn} \cd div_{\S} (\eta |_{\S}) \nonumber \\
&  &  + \12 \oints <\The |_{\S}, div_{\S} (\eta |_{\S}) \cd  g_o|_{\S}> 
 + \oint_{\S} \{ -\frac{1}{2}\The_{nn;n} 
+ \frac{1}{2}H_o \The_{nn} \} \cdot h  \nonumber \\ 
&  & + \oint_{\S} <\The |_{\S} , \tau> 
+ \12 \oints < (\na_n \The) |_{\S}, \Up |_{\S}> + \12
\oints (\na_n \The)_{nn} \cdot \Up_{nn}  \nonumber \\
& &  + \oints < (\na_n \The)(\vn, \cdot), \Up(\vn, \cdot)>
- \12 \oints < (\na_n \Up) |_{\S}, \The |_{\S}>  \nonumber \\
& & - \12 
\oints (\na_n \Up)_{nn} \cdot \The_{nn}  - \oints < (\na_n \Up)(\vn, \cdot), \The(\vn, \cdot)> \ ,
\end{eqnarray}
where each term on the right handside explicitly involves the free boundary data (\ref{freedata}).
Thus it follows from (\ref{bdrysim2}) 
that $\{ \Up, \eta, \tau, h \}$ satisfies the following boundary conditions on $\S$
\begin{equation} \label{bdrysim3}
\left\{ \begin{array} {ccl}
d_{\S} \eta_n + H_o \eta |_{\S} + \Pi_o( (\eta |_{\S})^*, \cd) + (\na_n \Up)(\vn, \cd) & = & 0 \\
- H_o \eta_n + \12 div_{\S}(\eta |_{\S}) + \12 H_o h - \12 \Up_{nn;n} & = & 0 
\\
\12 \eta_n - \12 h + \12 \Up_{nn} & = & 0 \\
\eta |_{\S} + \Up(\vn, \cd) & = & 0 \\
- \12 \eta_n g_o|_{\S} + \12 \Up |_{\S} & = & 0 \\ 
\eta_n \Pi_o - \mathcal{S}[\na_{\S} (\eta |_{\S})] + \12 div_{\S}(\eta |_{\S}) g_o|_{\S} + \tau - \12 (\na_n \Up)|_{\S} & = & 0 \ .
\end{array}
\right.
\end{equation}
On the other hand, by (\ref{divd}) and (\ref{divn}) we know that
\begin{equation}
\left\{ \begin{array} {rll}
(div \Up)_n & = & div_{\S} [\Up(\vn, \cd)] + \Up_{nn;n} - H_o \Up_{nn} + < \Up  |_{\S}, \Pi> \\
(div \Up )_{\d} & = & \Up_{n \d; n} + (div_{\S} [\Up |_{\S}] ) _{\d} - H_o \Up_{n \d} - g^{\a \b} \Up_{\a n}\Pi_{\d \b} \ .
\end{array}
\right.
\end{equation}
Hence, it is easily seen that (\ref{bdrysim3}) is equivalent 
to 
\begin{equation} \label{bdryrela1}
\left\{ \begin{array} {ccc}
div (\Up)_n & = & 0 \\
div (\Up)_{\d} & = & 0 \\
\Up |_{\S} & = & w g_o|_{\S} \\
h & = & w +  \Up_{nn} \\
\eta |_{\S} & = & - \Up(\vn, \cd) 
\end{array}
\right. 
\end{equation}
and
\begin{equation} \label{bdryrela2}
- \tau =  w \Pi + \mathcal{S}[\na_{\S} (\Up(\vn, \cd))] 
- \12 div_{\S}(\Up(\vn, \cd)) g_o|_{\S} - \12 (\na_n \Up)|_{\S}  \ ,
\end{equation}
where we replace $\eta_n$ be $w$. (Interesting simplification!)

So far our analysis has shown that
\begin{equation} \label{eqofad}
\left\{ \begin{array}{ccc}
         \triangle \vp - div(div\Up) & = & 0 \\
         \triangle \Up & = & 0 
        \end{array}
\right.  
\end{equation}
in $M$ and 
\begin{equation} \label{bdeqofad}
\left\{ \begin{array}{ccc}  
         \vp - \Up_{nn} & = & 0 \\
         \frac{\partial \vp}{\partial \vn} - div_{\S}\Up(\vn, \cdot) - (div \Up)_n  & = & 0 \\
         div \Up & = & 0 \\
         \Up |_{\S} & = & w g_o |_{\S} 
\end{array}
\right. 
\end{equation}
on $\S$, where $w$ is a parameter function. 
Now it follows from (\ref{eqofad}) that $\triangle (div \Up) = 0,$  thus integrating by parts and using the fact that
$$(div \Up)_i = O(r^{-\del -2}), \ \  D_j(div \Up_i) = O(r^{-\del-3}) \ \mathrm{and} \ \ \del > -\frac{3}{2} $$
we see that $div \Up \equiv 0$ in $M$.
Therefore, (\ref{eqofad}) and (\ref{bdeqofad}) become
\begin{equation} \label{seqofad}
\left\{ \begin{array}{ccc}
         \triangle \vp & = & 0 \\
         \triangle \Up & = & 0 
        \end{array}
\right. \hspace{1cm} in \ M
\end{equation}
and 
\begin{equation} \label{sbdeqofad}
\left\{ \begin{array}{ccc}  
         \vp - \Up_{nn} & = & 0 \\
         \frac{\partial \vp}{\partial \vn} - div_{\S}\Up(\vn, \cdot) & = & 0 \\
         div \Up & = & 0 \\
         \Up |_{\S} & = & w g_o |_{\S} 
\end{array}
\right. \hspace{.3cm} on \ \S \ ,
\end{equation}
which proves Lemma \ref{obstruction}.  
\end{proof}

It is easily seen that $(\Up, \vp) = (g_o, 1)$ satisfies both 
(\ref{seqofad}) and (\ref{sbdeqofad}). To eliminate such a trivial solution, 
we choose $\del \in (-1, -\frac{1}{2}]$ throughout the rest of our discussion. The following criteria now follows directly from the Implicit Function Theorem and our analysis above. (We note that $T$ has complemented kernel because its kernel is of finite dimension.)

\begin{proposition}
Let $\del \in (-1, -\frac{1}{2}]$ and $k - \frac{3}{2} > 3$, if (\ref{seqofad}) and (\ref{sbdeqofad}) only admit zero solution of $(\Up, \vp) \in \mathcal{H}^{2-k}_{-\del-1, M} \times H^{2-k}_{-\del-1, M}$, then there exists a neighborhood $U$ of $g_o$ in $\Mepo$ and a neighborhood $V$ of $g_o$ in $\Mepm$ so that, for any $\gi \in U$, 
there exists a scalar flat and static metric $\go \in V$ satisfying 
the geometric boundary condition \em{({\bf bd})}.
\end{proposition}

\section{Description of Coker(T) in Case $\Om = B_1$}
\label{cokernel}
From now on, we concentrate on the important case $(\Om, \S) = (B_1, S^2)$ and we will obtain an explicit description of the cokernel of $T$. First, we claim that $(\ref{seqofad})$ and $(\ref{sbdeqofad})$ admits no non-trivial rotationally symmetric solutions. To see that, let $(\Up, \vp)$ be such a solution with the form
\begin{equation}
\left\{ 
\begin{array}{ccc}
\Up_i(\partial_r, \partial_r) & = & a(r) \\
\Up_i(\partial_r, \cdot)|_{\partial{B_r}} &  = & 0 \\
\Up_i(\cdot, \cdot) |_{\partial{B_r}} & = & d(r) \cdot r^2 g_o|_{S^2} \\
\vp(x) & = & \vp(r)
\end{array}
\right.
\end{equation} 
where $r = |x|$ and $a(r), d(r), \vp(r)$ is a single variable function of $r$. The fact that $\vp$ is harmonic directly implies that that $\vp(r)=0$ because of the boundary condition and the decay assumption at $\infty$. Thus $(\ref{seqofad})$ and $(\ref{sbdeqofad})$ are reduced to a coupled ODEs
\begin{equation} \label{0ode}
\left\{
\begin{array}{ccc}
a^{\pr \pr}(r) + \frac{2}{r}a^{\pr}(r) - \frac{4}{r^2}[a(r) - d(r)] & = & 0 \\
d^{\pr \pr}(r) + \frac{2}{r}d^{\pr}(r) + \frac{2}{r^2}[a(r) - d(r)] & = & 0 
\end{array}
\right.
\end{equation}
with the boundary condition
\begin{equation} \label{0bdry}
\left\{
\begin{array}{ccc}
a(1) & = & 0 \\
a^{\pr}(1) -2 d(1) & = & 0 
\end{array}
\right. \ . 
\end{equation}
It follows from $(\ref{0ode})$ that  
\begin{equation}
r^3 a^{(4)}(r) + 8r^2 a^{(3)}(r) + 8ra^{\pr \pr}(r) - 8a^{\pr}(r) = 0 \ ,
\end{equation}
which, together with the decay assumption, shows that
\begin{equation}
\left\{
\begin{array}{ccc}
a(r) & = & B r^{-1} + C r^{-3} \\
d(r) & = & B r^{-1} - \12 C r^{-3} \ .
\end{array}
\right.
\end{equation}
It follows from $(\ref{0bdry})$ that both $B$ and $C$ are $0$.

Next, we follow the separation of variable method employed by Regge and Wheeler in \cite{Regge_Wheeler} and also by Hu in \cite{Hu} to decompose the tensor $\Up$ and the function $\vp$ using tensor harmonics. 
Keeping the same notation as in \cite{Regge_Wheeler}, we let 
$$\{ Y^M_L(\theta, \b) \ | \ M = 1, 2, \ldots, M_L \}$$ 
denote the set of spherical harmonics of degree $L = 1, 2, 3, \dots $, where $M_L$ is the dimension of the space of homogeneous harmonic polynomials in $\R^3$. Since  (\ref{seqofad}) and (\ref{sbdeqofad}) admit no non-trivial rotationally symmetric solutions, it suffices for us to look for solutions of the following two types: \\
\noindent \underline{Type (I)}:
\begin{equation}
L \geq 2: \ 
\left\{ 
\begin{array}{ccc}
\Up_M(\partial_r, \partial_r) & = & a(r) \cd \yml \\
\Up_M(\partial_r, \cdot)|_{\partial{B_r}} &  = & b(r) \cd d_{S^2} \yml \\
\Up_M(\cdot, \cdot) |_{\partial{B_r}} & = & r^2 [ c(r) \cd Hess_{S^2} \yml \\
& &  + d(r) \cd \yml g_o|_{S^2}] \\
\vp_M & = & c_o \frac{1}{r^{(L+1)}} \yml 
\end{array}
\right. 
\end{equation}
\begin{equation}
L = 1: \ \ \ \  
\left\{ 
\begin{array}{ccc}
\Up_M(\partial_r, \partial_r) & = & a(r) \cd Y^M_1 \\
\Up_M(\partial_r, \cdot)|_{\partial{B_r}} &  = & b(r) \cd d_{S^2} Y^M_1 \\
\Up_M(\cdot, \cdot) |_{\partial{B_r}} & = & r^2 d(r) \cd Y^M_1 g|_{S^2} \\
\vp_M & = & c_o \frac{1}{r^{2}} Y^M_1 
\end{array}
\right.
\end{equation} 
\noindent \underline{Type (II)}:
\begin{equation}
L \geq 2: \ 
\left\{ 
\begin{array}{ccc}
\hat{\Up}_M(\partial_r, \partial_r) & = & 0 \\
\hat{\Up}_M(\partial_r, \cdot)|_{\partial{B_r}} &  = & b(r) \cd ( d_{S^2} \yml )^* \\
\hat{\Up}_M(\cdot, \cdot) |_{\partial{B_r}} & = &  c(r) \cd (Hess_{S^2} \yml)^* \\
\hat{\vp}_M & = & 0
\end{array}
\right.
\end{equation}  
\begin{equation}
L = 1: \ 
\left\{ 
\begin{array}{ccc}
\hat{\Up}_M(\partial_r, \partial_r) & = & 0 \\
\hat{\Up}_M(\partial_r, \cdot)|_{\partial{B_r}} &  = & b(r) \cd ( d_{S^2} \yml )^* \\
\hat{\Up}_M(\cdot, \cdot) |_{\partial{B_r}} & = &  0 \\
\hat{\vp}_M & = & 0
\end{array}
\right.
\end{equation}  
where $a(r), b(r) , c(r)$ and $d(r)$ are single variable functions of $r$, $c_o$ is a constant, $d_{S^2}\yml$ and $Hess_{S^2}\yml$ represent the exterior derivative of $\yml$ and the Hessian of $\yml$ on $S^2$, $(d_{S^2}\yml)^*$ 
and $(Hess_{S^2}\yml)^*$ 
are defined to be the dual of $d_{S^2}\yml$ and $Hess_{S^2}\yml$ in the following sense:
\begin{equation}
\begin{array}{c}
(d_{S^2}\yml)^*_{\d}  =  \ep^{\l}_{\d} \cd  ({d_{S^2}Y^M_L})_{\l }  
 \\
(Hess_{S^2}\yml)^*_{\a \d}  =  \12 \{ \ep^{\l}_{\a} \cd (Hess_{S^2} \yml)_{\l \d} + \ep^{\l}_{\d} \cd (Hess_{S^2} \yml)_{\l \a} \} \ ,
\end{array}
\end{equation}
where $\ep^{\l}_{\d}$ is a $(1, 1)$ tensor on $S^2$ defined by 
\begin{equation}
\begin{array}{cclc}
\ep^{\theta}_{\theta}  =  0, \ & \ep^{\b}_{\theta}  =   - \frac{1}{\sin\theta} \\
\ep^{\b}_{\b} =  0, \ & \ep^{\theta}_{\b} = \sin(\theta)
\end{array}
\end{equation}
in the standard spherical coordinates on $S^2$ (we note that $g_o|_{S^2} = (d\theta)^2 + (\sin(\theta)d\b)^2 $). 
It is easily seen that $\ep$ is a linear isometry of $T(S^2)$ which rotates every tangent vector $\frac{\pi}{2}$ clockwise. In particular, $\ep$ is parallel, i.e. $\na_{S^2} \ep = 0$.

First we look for Type (I) solutions. Straightforward calculation, though not quite a pleasant thing to do, shows that $(\ref{seqofad})$ and $(\ref{sbdeqofad})$ are reduced to the following system of coupled ODEs,
\begin{equation}
\left\{ 
\begin{array}{ccc} \label{odesys2}
d^{\pr \pr}(r) + \frac{2}{r} d^{\pr}(r) - \frac{2}{r^2}d(r) + \frac{2}{r^2} a(r) - \frac{L(L+1)}{r^2} [ d(r) - 2c(r) ]  & = & 0 \\
c^{\pr \pr}(r) + \frac{2}{r} c^{\pr}(r) + \frac{2}{r^2} c(r) + \frac{4}{r^3}b(r) -\frac{L(L+1)}{r^2} c(r) & = & 0 \\  
b^{\pr \pr}(r) - \frac{4}{r^2} b(r) + \frac{2}{r}a(r) -  \frac{2}{r}d(r) - 
\frac{2}{r}c(r) - \frac{L(L+1)}{r^2} [ b(r) - 2rc(r) ] & = & 0 \\
a^{\pr \pr}(r) + \frac{2}{r} a^{\pr}(r) - \frac{4}{r^2}a(r) + \frac{4}{r^2}d(r) -
 \frac{L(L+1)}{r^2} [ a(r) - \frac{4}{r}b(r) + 2c(r) ] & = & 0
\end{array}
\right.
\end{equation}
with the boundary condition
\begin{equation} \label{odebdry2}
\left\{
\begin{array}{ccc}
c_o - a(1) & = & 0 \\
c_o - L b(1) & = & 0 \\
a^{\pr}(1) + 2a(1) - 2d(1) - L(L+1)b(1) & = & 0 \\
b^{\pr}(1) + 2b(1) + d(1) & = & 0 \\
c(1) & = & 0 
\end{array}
\right.
\end{equation}
for $L \geq 2$, and
\begin{equation}
\left\{ 
\begin{array}{ccc} \label{odesys1}
r^2 d^{\pr \pr}(r) + 2r d^{\pr}(r) - 4 d(r) + 2a(r) - \frac{4}{r}b(r) & = & 0 \\
r b^{\pr \pr}(r) - \frac{6}{r}b(r) + 2a(r) - 2d(r) & = & 0 \\
r^2 a^{\pr \pr}(r) + 2r a^{\pr}(r) - 6a(r) + 4d(r) + \frac{8}{r} b(r) & = & 0
\end{array}
\right.
\end{equation}
with the boundary condition
\begin{equation} \label{odebdry1}
\left\{
\begin{array}{ccc}
c_o - a(1) & = & 0 \\
c_o - b(1) & = & 0  \\
a^{\pr}(1) + 2a(1) - 2d(1) - 2 b(1)  & = & 0 \\
b^{\pr}(1) + 2b(1) + d(1)  & = & 0 
\end{array}
\right.
\end{equation}
for $L =1$.
When $L \geq 2$, it follows from (\ref{odesys2}) and plain calculation that
\begin{eqnarray}
0 & = & r^5 \cd  c^{(5)}(r) + 16r^4 \cd c^{(4)}(r) + [72 - 2L(L+1)]r^3 \cd c^{(3)}(r)  \nonumber \\
& &  + [ 96 - 12 L(L+1)]r^2 \cd c^{\pr \pr}(r)  \nonumber \\
& &  + [ L^2(L+1)^2 - 14 L(L+1) + 24]r \cd c^{\pr}(r) 
\end{eqnarray}
which shows that 
\begin{equation}
c(r) = A r^{-L-3} + Br^{-L-1} + C  + D r^{L-2} + E r^L \ .
\end{equation}
The decay assumption on $\Up$ near $\infty$ implies that 
\begin{equation}
c(r) = A r^{-L-3} + Br^{-L-1} \ .
\end{equation}
It is easily checked that the boundary condition (\ref{odebdry2}) is sufficient to force both $A$ and $B$ to vanish, hence yields that $c(r) = 0$. Then it follows from (\ref{odesys2}) and (\ref{odebdry2}) that $a(r), b(r)$ and $d(r)$ all vanish identically.
When $L =1$, (\ref{odesys1}) implies that 
\begin{equation}
r^3 b^{\pr \pr \pr}(r) + 5r^2 b^{\pr \pr}(r) - 2r b^{\pr}(r) - 6b(r) = 0,
\end{equation}
which gives that
\begin{equation}
b(r) = A r^{-1} + Br^{-3} \ .
\end{equation}
Now it can be checked that (\ref{odebdry1}) is not sufficient to force both $A$ and $B$ to vanish. Indeed, we have $B=0$ and $A$ can be any number. Hence, the solutions space is spanned by 
\begin{equation}
(a(r), b(r), d(r)) = ( \frac{1}{r^2}, \frac{1}{r}, -\frac{1}{r^2}) \ . 
\end{equation}

Next we turn to Type (II) solutions. Similar calculation reveals that we have a system of coupled ODEs
\begin{equation} \label{odesysodd2}
\left\{
\begin{array}{ccc}
c^{\pr \pr}(r) - \frac{2}{r}c^{\pr}(r) + (4 - L^2 -L ) \frac{1}{r^2} c(r) 
+ \frac{4}{r}b(r) & = & 0 \\
b^{\pr \pr}(r) - (4 + L^2 + L ) \frac{1}{r^2} b(r) + (-2 + L^2 + L)
 \frac{1}{r^3}c(r) & = & 0
\end{array}
\right.
\end{equation}
with the boundary condition
\begin{equation} \label{odebdryodd2}
\left\{
\begin{array} {rcc}
c(1) & = & 0 \\
b^{\pr}(1) & = & -2 b(1)  
\end{array}
\right.
\end{equation}
for $ L \geq 2$, and 
\begin{equation} \label{odesysodd1}
b^{\pr \pr}(r) - (4 + L^2 + L ) \frac{1}{r^2} b(r) = 0
\end{equation}
with the boundary condition
\begin{equation} \label{odebdryodd1}
b^{\pr}(1)  = -2 b(1)  
\end{equation}
for $ L = 1 $, where we use the fact that 
\begin{equation}
div_{S^2} [(d_{S^2} \yml)^*]  = 0 \mathrm{ \ and\ } tr_{g|_{S^2}} [(Hess_{S^2}\yml)^*] = 0 \ . 
\end{equation}
When $L \geq 2$, it follows from (\ref{odesysodd2}) that 
\begin{eqnarray}
0 & = & r^4 \cd c^{(4)}(r) - 2L(L+1) r^2 \cd c^{\pr \pr}(r)  + 4L(L+1)r \cd c^{\pr}(r) \nonumber \\
& &+ [L^2(L+1)^2 - 6L(L+1)] \cd c(r) \ ,
\end{eqnarray}
which gives that
\begin{equation}
c(r) = A r^{-L} + Br^{2-L} + Cr^{L+1} + Dr^{L+3} \ .
\end{equation}
Since $r^{-2} \cd c(r)$ decays at $\infty$, we have that
\begin{equation}
c(r) = A r^{-L} + Br^{2-L}  \ .
\end{equation}
It is readily seen that (\ref{odebdryodd2}) forces both $A$ and $B$ to vanish, hence $C(r) = 0$ and $b(r) = 0$. 
When $L = 1$, (\ref{odesysodd1}) directly gives that $b(r) = Ar^{-2} + Br^{3}$, which together with the decay and boundary condition shows that 
\begin{equation}
b(r) = A r^{-2}
\end{equation}
is the only solution. 

To summarize our analysis, we first replace the notation $Y^M_i$ by $\xi_i(\theta, \beta)$ for $i=1,2,3$ and define
\begin{equation}
\left\{ 
\begin{array}{ccc}
\Up_i(\partial_r, \partial_r) & = & r^{-2}\xi_i(\theta, \beta) \\
\Up_i(\partial_r, \cdot)|_{\partial{B_r}} &  = & r^{-1}d_{S^2}\xi_i(\theta, \beta) \\
\Up_i(\cdot, \cdot) |_{\partial{B_r}} & = & -r^{-2}\xi_i(\theta, \beta)(g_o|_{\partial B_r}) \\
\vp_i & = & r^{-2}\xi_i(\theta, \beta)  
\end{array}
\right.
\end{equation}
and
\begin{equation}
\left\{ 
\begin{array}{ccl}
\hUp_i(\partial_r, \partial_r) & = & 0 \\
\hUp_i(\partial_r, \cdot)|_{\partial{B_r}} &  = & r^{-2} (d_{S^2}\xi_i(\theta, \beta))^* \\
\hUp_i(\cdot, \cdot) |_{\partial{B_r}} & = & 0 \\
\hvp_i & = & 0 \ \ \ \ .
\end{array}
\right. 
\end{equation}
Our calculation above then shows that the solution space of (\ref{seqofad}) 
and (\ref{sbdeqofad}) is spanned by
\begin{equation} \label{solu8283}
\{ (\Up_i, \vp_i), (\hUp_i, \hvp_i) \ | \ i = 1, 2, 3 \} \ .
\end{equation}
The following characterization of the cokernel of $T$ and the image of $T$ now follow directly from (\ref{solu8283}), (\ref{bdryrela1}), (\ref{bdryrela2}) and the general
linear elliptic theory \cite{McOwen}.
\begin{proposition} \label{image}
If $\Om = B_1$, then 
 $$Coker(T) = span\{ (\Up_i, \vp_i, \eta_i, 0, 0 ),   (\hUp_i, 0, \heta_i, 0, 0 )\ | \ i=1, \ 2, \ 3\} $$
where 
\begin{equation}
\eta_i =  - d_{S^2}\xi_i(\theta, \phi) - \xi_i(\theta,\phi)dr
\end{equation}
and  
\begin{equation}
\heta_i =  - ( d_{S^2}\xi_i(\theta, \phi) )^* \ .
\end{equation} 
Furthermore,
given $(\Psi, \psi, \zeta, \sigma, \tilde{h}) \in \mathcal{H}^{k-2}_{\del-2, M} \times H^{k-2}_{\del-2, M} \times \mathcal{H}^{k - \frac{3}{2}}_{(\S, \R^3)} \times \mathcal{H}^{k - \frac{1}{2}}_{\S} \times
   H^{k - \frac{3}{2}}_{\S} $,  
\begin{equation} \label{fulleq}
 T(\The, \phi) = 
 \left( 
 \begin{array}{c}
 -\frac{1}{2} \triangle \The - Hess \phi \\
 \triangle \phi \\
 div \The - \frac{1}{2} d (tr \The) \\
 \The |_{\S} \\
 -\frac{1}{2}\The_{nn;n} + \frac{1}{2}H_0\The_{nn}
 \end{array}
 \right)
 =
  \left( 
 \begin{array}{c}
 \Psi \\
 \psi \\
 \zeta \\
 \sigma \\
 \tilde{h}
 \end{array}
 \right)
\end{equation}
has a solution $ (\The, \phi) \in \mathcal{H}^{k}_{\del, M} \times H^k_{\del, M}$ 
if and only if 
\begin{equation} \label{imconst}
\left\{ 
\begin{array}{ccc}
\int_{\R^3\setminus B_1} < \Psi, \Up_i > +  
 \int_{\R^3 \setminus B_1} <\psi, \vp_i> +
 \oint_{S^2} < \omega, \eta_i> & = & 0 \\
\int_{\R^3\setminus B_1} < \Psi, \hUp_i > + \int_{\R^3 \setminus B_1} <\psi, \hvp_i> +  
 \oint_{S^2} < \omega, \heta_i> & = & 0 
\end{array}
\right.
\end{equation}
for all $i \in \{ 1, 2, 3\}$.
Hence, $ (\Psi, \psi, \zeta, \sigma, 
\tilde{h}) \in Image(D^{\prime}\Phi_o$) if and only if
(\ref{imconst}) holds.
\end{proposition}

\section{Proof of the Main Theorem}
\label{pfofthm}
We prove our main theorem based on the following basic observation.

\vspace{.2cm}
\noindent \underline{Fact}: \emph{For any $i \in \{1, 2, 3 \}$, $(\Up_i, \vp_i, \eta_i)$ is ``odd'' under the reflection about the coordinate plane not containing $e_i$ while $(\hUp_i, \hvp_i, \heta_i)$ is ``odd'' under the reflection about the coordinate planes containing $e_i$. Hence, $(\ref{imconst})$ holds automatically if $(\Psi, \psi, \zeta, \sigma, \tilde{h})$ is ``even''(or invariant) under reflections about all the coordinate planes. }
\vspace{.2cm}

Keeping this in mind, we define $G$ to be the finite group of isometries of 
$\R^3$ that is generated by all reflections with respect to coordinate planes. It is easily seen that $G$ is isomorphic to $\Z_2 \times \Z_2 \times \Z_2$.

\begin{definition}
$\Mepme, \Mepoe, \Fepe, \mathcal{H}^{l,G}_{\del, M}, 
H^{l,G}_{\del, M}, \mathcal{H}^{l,G}_{(S^2, \R^3)}, \mathcal{H}^{l,G}_{S^2},
H^{l,G}_{S^2}$ is defined to be the $G$-invariant subspace of  
$\Mepm, \Mepo, \Fep, \mathcal{H}^l_{\del, M}, 
H^l_{\del, M}, \mathcal{H}^l_{(S^2, \R^3)}, \mathcal{H}^l_{S^2},
H^l_{S^2}$.
\end{definition}

The fact that $G$ consists of isometries of $\R^3$ implies that
\begin{equation} \label{simofphi}
\iota^*( \Phi(\go, f, \gi)) = \Phi(\iota^*(\go, f, \gi)), \ \forall 
\iota \in G.
\end{equation}
Hence, we have a well defined map ${\Phi}^G$ which is the restriction of $\Phi$ to the $G$-invariant subspaces,
$$ {\Phi}^G: \Mepme \times \Fepe \times \Mepoe 
\longrightarrow
\mathcal{H}^{k-2, G}_{\del-2, M} \times
H^{k-2,G}_{\del-2, M} \times
\mathcal{H}^{k-\frac{3}{2}, G}_{(S^2, \R^3)} \times 
\mathcal{H}^{k-\frac{1}{2},G}_{S^2} \times
H^{k-\frac{3}{2},G}_{S^2}.$$
We let $D^\prime \Phi^G_o$ denote the partial differential of $\Phi^G$ at $(g_o, 1, g_o)$ with respect to the first two arguments.

\begin{proposition}
$$ D^\prime \Phi^G_o: 
\mathcal{H}^{k,G}_{\del, M} \times H^{k,G}_{\del, M} 
\longrightarrow
\mathcal{H}^{k-2,G}_{\del-2, M} \times
H^{k-2,G}_{\del-2, M} \times
\mathcal{H}^{k-\frac{3}{2},G}_{(S^2, \R^3)} \times 
\mathcal{H}^{k-\frac{1}{2},G}_{S^2} \times
H^{k-\frac{3}{2},G}_{S^2}$$
is a surjective map.
\end{proposition}

\begin{proof}
Let $(\Psi, \psi, \zeta, \sigma, \tilde{h})$ be any element in the target 
space. By definition we have that
\begin{equation} \label{simofim}
\iota^*{(\Psi, \psi, \zeta, \sigma, \tilde{h})} = (\Psi, \psi, \zeta, \sigma, \tilde{h}), \ \forall \iota \in G.
\end{equation}
Proposition \ref{image} implies that $ \exists \ (\Gam, \phi) \in
\mathcal{H}^k_{\del, M} \times H^{k}_{\del, M}$ so that
\begin{equation} \label{firstsol}
D^\prime\Phi^G_o(\Gam, \phi) = 
(\Psi, \psi, \zeta, \sigma, \tilde{h}).
\end{equation}
On the other hand, (\ref{simofphi}) gives that
\begin{equation}
D^\prime \Phi^G_o (\iota^*(\Gam), \iota^*(\phi)) =
\iota^*(\Psi, \psi, \zeta, \sigma, \tilde{h}). 
\end{equation}
Hence, (\ref{simofim}) implies that 
\begin{equation}
D^\prime \Phi^G_o (\iota^*(\Gam), \iota^*(\phi)) =
(\Psi, \psi, \zeta, \sigma, \tilde{h}),
\end{equation}
which, together with (\ref{firstsol}), gives that
\begin{equation}
D^\prime \Phi^G_o (\frac{1}{8} \sum_{\iota \in G} \iota^*(\Gam), 
\frac{1}{8} \sum_{\iota \in G} \iota^*(\phi) ) =
(\Psi, \psi, \zeta, \sigma, \tilde{h}).
\end{equation}
Since $ (\frac{1}{8} \sum_{\iota \in G} \iota^*(\Gam), 
\frac{1}{8} \sum_{\iota \in G} \iota^*(\phi)) \in \mathcal{H}^{k,G}_{\del, M} 
\times H^{k,G}_{\del, M}$, we conclude that
$$ D^\prime \Phi^G_o(\frac{1}{8} \sum_{\iota \in G} \iota^*(\Gam), \frac{1}{8} \sum_{\iota \in G} \iota^*(\phi) ) = (\Psi, \psi, \zeta, \sigma, \tilde{h}),
$$
which shows the subjectivity of $ D^\prime \Phi_o $. 
\end{proof}

Our main existence theorem now follows readily from the above proposition and 
the Inverse Function Theorem.

\vspace{.3cm}

\noindent \emph{Acknowledgment} \ 
I am very grateful to my Ph.D. advisor Professor Richard Schoen, who suggested this problem and gives me constant directions and encouragements. I also would like to thank Professor Robert Bartnik and Professor Hubert Bray for many stimulating discussions during their visit at the AIM-Stanford workshop on General Relativity in April 2002. Finally, I would like to thank Professor Vincent Moncrief for explaining to me the work of \cite{Regge_Wheeler}. 

\bibliographystyle{plain}
\bibliography{static_ext}

\end{document}